# Organization of the bacterial nucleoid by DNA-bridging proteins and globular crowders.

**Marc JOYEUX[1*]**

[1]Laboratoire Interdisciplinaire de Physique, CNRS and Université Grenoble Alpes, St Martin d'Hères, France

**\* Correspondence:**
Marc JOYEUX
marc.joyeux@univ-grenoble-alpes.fr

## Abstract

The genomic DNA of bacteria occupies only a fraction of the cell called the nucleoid, although it is not bounded by any membrane and would occupy a volume hundreds of times larger than the cell in the absence of constraints. The two most important contributions to the compaction of the DNA coil are the cross-linking of the DNA by nucleoid proteins (like H-NS and StpA) and the demixing of DNA and other abundant globular macromolecules which do not bind to the DNA (like ribosomes). The present work deals with the interplay of DNA-bridging proteins and globular macromolecular crowders, with the goal of determining the extent to which they collaborate in organizing the nucleoid. In order to answer this question, a coarse-grained model was developed and its properties were investigated through Brownian dynamics simulations. These simulations reveal that the radius of gyration of the DNA coil decreases linearly with the effective volume ratio of globular crowders and the number of DNA bridges formed by nucleoid proteins in the whole range of physiological values. Moreover, simulations highlight the fact that the number of DNA bridges formed by nucleoid proteins depends crucially on their ability to self-associate (oligomerize). An explanation for this result is proposed in terms of the mean distance between DNA segments and the capacity of proteins to maintain DNA bridging in spite of the thermal fluctuations of the DNA network. Finally, simulations indicate that non-associating proteins preserve a high mobility inside the nucleoid while contributing to its compaction, leading to a DNA/protein complex which looks like a liquid droplet.

In contrast, self-associating proteins form a little deformable network which cross-links the DNA chain, with the consequence that the DNA/protein complex looks more like a gel.

## 1 Introduction

The contour length of the genomic DNA of *E. coli* bacteria is of the order of $L \approx 1.5$ mm and its persistence length $\xi \approx 50$ nm. In the absence of any constraint, the DNA would form a coil with radius of gyration $R_g = \sqrt{L\xi/6} \approx 3.5$ μm (Teraoka, 2002). This radius is significantly larger than the dimensions of *E. coli* cells, which have the shape of an approximate sphero-cylinder with average diameter ≈1 μm, average length ≈2 μm, and average volume ≈1.3 μm³. Since the DNA of prokaryotes is not bounded by any membrane, except for the cell membrane itself, one could easily imagine that the DNA spreads throughout the cell. Instead, the chromosomic DNA of bacteria occupies in most cases only a fraction of the cell, which is called the nucleoid and is composed essentially of DNA (≈80%), RNA (≈10%) and a dozen of relatively abundant proteins (≈10%) (Stonington and Pettijohn, 1971; Azam et al., 1999; Sherratt, 2003). The volume of the nucleoid depends sensitively on several factors, such as the richness of nutrients (Yazdi et al., 2012; Endesfelder et al., 2013; Jin et al., 2013; Jin et al., 2015; Stracy et al., 2015), the cell cycle step (Fisher et al., 2013; Spahn et al., 2014), and the eventual treatment with antibiotics (Cabrera et al., 2009; Bakshi et al., 2012; Jin et al., 2013; Cagliero and Jin, 2013; Bakshi et al., 2014; Jin et al., 2015; Stracy et al., 2015; Bakshi et al., 2015), but the nucleoid fills quite rarely the whole cell. The mechanism leading to the compaction of the DNA inside the nucleoid is a longstanding but still lively debated question (Zimmerman, 2006; de Vries, 2010; Benza et al., 2012; Joyeux, 2015; Joyeux, 2016). The three most important contributions to the compaction of the DNA are presumably (a) the cross-linking of the DNA by DNA-bridging proteins like H-NS and StpA (Spassky et al., 1984; Spurio et al., 1992; Qin et al., 2019; Joyeux, 2021), (b) DNA supercoiling, that is, the winding about itself of the circular DNA in response to the torsional stress generated by topoisomerases (Cunha et al., 2001; Stuger et al., 2002; Bates and Maxwell, 2005), and (c) the demixing of DNA and other abundant globular macromolecules which do not bind to the DNA (presumably ribosomes), resulting in a segregative phase separation (Castelnovo and Gelbart, 2004; de Vries, 2006; Krotova et al., 2010; Zinchenko et al., 2014; Joyeux, 2016; Joyeux, 2018). It is tempting to study each of these contributions separately and to rely on the hypothesis that isolated contributions "sum up" in living cells. The additivity assumption is, however, a very questionable one. For example, it has recently been shown through coarse-grained modeling and Brownian dynamics simulations, that the effects of supercoiling and demixing are additive only in a limited range of biologically relevant values of the parameters, the reason being that the plectonemes generated by supercoiling are stiffer than torsionally relaxed DNA and consequently less easily compacted by globular macromolecules (Joyeux, 2019).

The present paper deals with the compaction of the genomic DNA in a medium which contains both DNA-bridging proteins and globular macromolecules that do not bind to the DNA (hereafter called "crowders"). The first point to ascertain is whether bridging proteins and globular crowders collaborate in compacting the bacterial DNA, or whether they do so only in a limited range of biologically relevant values, as is the case for supercoiling and crowders (Joyeux, 2019). This is an intriguing question from the physical point of view. Indeed, compaction of the DNA by bridging proteins proceeds via *associative* phase separation into a phase rich in both DNA and proteins and another phase deprived of both components. In contrast, demixing of DNA and globular crowders corresponds to a *segregative* phase separation into a phase enriched in DNA, but depleted in crowders, and another phase enriched in crowders, but depleted in DNA. Asking whether bridging proteins and globular crowders collaborate in compacting the bacterial DNA therefore amounts to determining whether the combination of associative and segregative phase separations is an efficient mechanism for organizing the nucleoid. Note that supercoiling is not taken into account in the present paper and that the organization of the bacterial nucleoid by DNA-bridging proteins, globular macromolecules and supercoiling is postponed to future work.

The second point explored in the present work deals with the eventual contribution of proteins' self-association to the compaction of the DNA coil by mixtures of DNA-bridging proteins and globular crowders. More precisely, it is known that some DNA-bridging proteins, in addition to binding to the DNA, are also able to self-associate (oligomerize). For example, the most abundant DNA-bridging nucleoid proteins, namely H-NS and StpA, are functional as dimers but can under certain conditions form larger oligomers (Smyth et al., 2000; Ceschini et al., 2000; Johansson et al., 2001; Badaut et al., 2002; Lim et al., 2012; Giangrossi et al., 2014; Boudreau et al., 2018). It has, moreover, been shown recently that proteins' self-association may increase dramatically their ability to shape the DNA coil (Joyeux, 2021). Therefore, the compaction of the DNA coil in mixtures of globular crowders and non-associating proteins will be compared to that of mixtures containing self-associating proteins, in order to quantify the role of proteins' self-association.

## 2    Model and simulations

## 2.1 Model

The models of DNA and self-associating DNA-bridging proteins have been described in detail in the Supporting Material of (Joyeux, 2021) and are summarized below for the sake of clarity The model of globular crowders, as well as the minor adjustment made to (Joyeux, 2021) to model non-associating proteins, are next described in detail.

As is illustrated in Fig. 1, the system is composed of one long circular DNA chain, a number $P = 0$, 50, 100, 200, or 400 of short linear protein chains, and a number $C = 0$, 500, 900, 1200, 1500, 1750, or 1875 of spherical crowders, which are all enclosed in a confinement sphere of radius $R_0 = 120$ nm. Concentrations of nucleotides, proteins and crowders are of the same order of magnitude as *in vivo* ones. The DNA chain and the protein chains are composed of beads of radius $a = 1.0$ nm, which are separated at equilibrium by a distance $l_0 = 2.5$ nm and are connected by springs. Each DNA bead represents 7.5 base pairs (bp) and the circular chain contains $n = 2880$ beads, equivalent to 21600 bp. As is sketched in Fig. 2, each protein chain is composed of 7 beads with index $m$ ($1 \leq m \leq 7$). Terminal beads $m = 1$ and 7 represent DNA-binding sites. For self-associating protein chains, beads $m = 2$ and 6 represent oligomerization sites, according to Model I of (Joyeux, 2021). Finally, crowders are modeled as spheres of radius $b = 7.4$ nm.

The total potential energy of the system, $E_{\text{pot}}$, is the sum

$$E_{\text{pot}} = E^{\text{DNA}} + \sum_{j=1}^{P} E^j + \sum_{j=1}^{P} E^{\text{DNA}/j} + \sum_{j=1}^{P-1} \sum_{J=j+1}^{P} E^{j/J} + E^* , \tag{1}$$

where $E^{\text{DNA}}$ is the internal energy of the DNA chain (Eqs. (S1)-(S7) of (Joyeux, 2021)), $E^j$ the internal energy of protein chain $j$ (Eqs. (S8)-(S13) of (Joyeux, 2021)), $E^{\text{DNA}/j}$ the interaction energy of the DNA chain and protein chain $j$ (Eqs. (S14)-(S17) of (Joyeux, 2021)), and $E^{j/J}$ the interaction energy of protein chains $j$ and $J$ (Eqs. (S18)-(S22) of (Joyeux, 2021)). Several points are worth mentioning concerning these terms :

- the DNA bending force constant (Eq. (S3) of (Joyeux, 2021)) was chosen so that the model reproduces the known persistence length of DNA under usual conditions (50 nm),

- the two terminal beads of each protein chain ($m = 1$ and 7) rotate freely around beads $m = 2$ and 6, respectively (Eq. (S10) of (Joyeux, 2021)), so that protein chains are significantly less rigid than the DNA chain, as is usually the case *in vivo*,

- beads $m = 1$ and 7 can bind to the DNA chain with a maximum binding energy of $7.8\,k_{\mathrm{B}}T$ (Eqs. (S15) and (S16) of (Joyeux, 2021)), which is comparable to the experimentally determined value of $10.0\,k_{\mathrm{B}}T$ (Ono et al., 2005),

- for self-associating proteins (Model I of (Joyeux, 2021)), beads $m = 2$ and 6 of one protein chain can bind to beads $m = 2$ and 6 of other protein chains (Eqs. (S19) and (S21) of (Joyeux, 2021)), so that protein chains spontaneously form clusters, as shown in the top-left vignette of Fig. 2 of (Joyeux, 2021). The binding interaction of oligomerization beads is modeled by a Lennard-Jones 3-6 potential of depth $\varepsilon_{\mathrm{LJ}}$ (Eq. (S19) of (Joyeux, 2021)), which was varied from $4\,k_{\mathrm{B}}T$ to $12\,k_{\mathrm{B}}T$ in (Joyeux, 2021). In the present work, $\varepsilon_{\mathrm{LJ}}$ was fixed to $10\,k_{\mathrm{B}}T$, close to the experimentally determined value of $10.2\,k_{\mathrm{B}}T$ for the enthalpy change upon forming a complex between two H-NS dimers (Ceschini et al., 2000).

The model of non-associating proteins used in the present work was derived from Model I of (Joyeux, 2021) by assuming that all beads of protein chain $j$ repel all beads of protein chain $J$, which amounts to replacing Eqs. (S18) to (S21) of (Joyeux, 2021) by

$$E^{j/J} = V_{\mathrm{ev}}^{j/J} = 8\,k_{\mathrm{B}}T \sum_{m=1}^{7} \sum_{M=1}^{7} F\big(\|\mathbf{R}_{jm} - \mathbf{R}_{JM}\|\,|r_0\big), \tag{2}$$

where $E^{j/J} = V_{\mathrm{ev}}^{j/J}$ denotes the repulsive interaction of protein chains $j$ and $J$. In Eq. (2), $\mathbf{R}_{jm}$ denotes the position of the center of bead $m$ of protein chain $j$, $r_0 = 3$ nm, and function $F\big((r|r_0)\big)$ is defined as in Eq. (S13) of (Joyeux, 2021)

$$\text{if} \leq r_0 : F\big((r|r_0)\big) = \frac{r_0^6}{r^6} - 2\frac{r_0^3}{r^3} + 1,$$

$$\text{if} > r_0 : F\big((r|r_0)\big) = 0 \,. \tag{3}$$

The last term in the right-hand side of Eq. (1) stands for the potential energy of the crowders. It is expressed in the form

$$E^* =$$
$$1000\,k_{\mathrm{B}}T \sum_{q=1}^{C} f\big(\|\mathbf{S}_q\|\big) + e_{\mathrm{DNA}}^2 \sum_{q=1}^{C} \sum_{k=1}^{n} H\big(\|\mathbf{S}_q - \mathbf{r}_k\| - a - b\big) +$$
$$e_{\mathrm{DNA}}^2 \sum_{q=1}^{C} \sum_{j=1}^{P} \sum_{m=1}^{7} H\big(\|\mathbf{S}_q - \mathbf{R}_{jm}\| - a - b\big) + e_{\mathrm{DNA}}^2 \sum_{q=1}^{C-1} \sum_{Q=q+1}^{C} H\big(\|\mathbf{S}_q - \mathbf{S}_Q\| - 2b\big) \,, \tag{4}$$

where $\mathbf{r}_k$ denotes the position of the center of bead $k$ of the DNA chain, $\mathbf{R}_{jm}$ the position of the center of bead $m$ of protein chain $j$, and $\mathbf{S}_q$ the position of the center of spherical crowder $q$. The first term in the right-hand side of Eq. (4) is the confinement term, which ensures that the centers of all spherical crowders remain inside the confinement sphere. Function $f(r)$ is defined as in Eq. (S7) of (Joyeux, 2021) if $r \leq R_0 : f(r) = 0$,

if $r > R_0 : f(r) = \frac{r^6}{R_0^6} - 1$ . (5)

The last three terms in the right-hand side of Eq. (4) describe the repulsive interaction between a spherical crowder and a DNA bead, a spherical crowder and a protein bead, and two spherical crowders, respectively. All these interactions are modeled by repulsive Debye-Hückel potentials with hard core, where function $H(r)$ is defined as in Eq. (S5) of (Joyeux, 2021)

$H(r) = \frac{1}{4\pi\varepsilon r} \exp\left(-\frac{r}{r_D}\right)$ . (6)

In Eq. (6), $\epsilon = 80\epsilon_0$ denotes the dielectric constant of the buffer and $r_D = 1.07$ nm its Debye length, whose value corresponds to a concentration of monovalent salts of 100 mM. For the sake of simplicity, it was assumed that the strength of the repulsion is identical for all particles and corresponds to the case where a charge $e_{DNA} = -3.525\bar{e}$, with $\bar{e}$ the absolute charge of the electron, is placed at the center of each particle (Joyeux, 2021).

It was shown in (Joyeux, 2017) that the compaction of the DNA chain by spherical crowders is governed by the effective volume fraction of crowders

$\rho = C \left(\frac{b + \Delta b}{R_0}\right)^3$ , (7)

where $2(b + \Delta b)$ is the distance at which the repulsion energy between two spherical crowders is equal to $k_B T$. For the repulsive potential in Eq. (4), one gets $\Delta b = 0.865$ nm. Simulations performed with $C = 0$, 500, 1000, 1200, 1500, 1750 and 1875 crowders therefore correspond to $\rho = 0$, 0.16, 0.33, 0.39, 0.49, 0.57 and 0.61, respectively. The reader is reminded that hard spheres become jammed (i.e. they can no longer move because they are too tightly packed) at a volume fraction close to 0.66 (Chaudhuri et al., 2010).

## 2.2   Simulations

The dynamics of the model was investigated by integrating numerically overdamped Langevin equations. Practically, the new position vector for each particle (DNA bead, protein bead, or spherical crowder), $\mathbf{x}_j^{(i+1)}$, was computed from the current position vector, $\mathbf{x}_j^{(i)}$, according to

$$\mathbf{x}_j^{(i+1)} = \mathbf{x}_j^{(i)} + \frac{D_j}{k_\mathrm{B}T}\Delta t\, \mathbf{F}_j^{(i)} + \sqrt{2D_j\,\Delta t}\,\boldsymbol{\xi}_j^{(i)} , \qquad (8)$$

where the translational diffusion coefficient $D_j$ is equal to $(k_\mathrm{B}T)/(6\pi\eta a)$ for DNA and protein beads and to $(k_\mathrm{B}T)/(6\pi\eta b)$ for spherical crowders. $\eta = 0.00089$ Pa.s is the viscosity of the buffer at $T = 298$ K. $\mathbf{F}_j^{(i)}$is the vector of inter-particle forces arising from the potential energy $E_\mathrm{pot}$ and $\boldsymbol{\xi}_j^{(i)}$ a vector of random numbers extracted at each step $i$ from a Gaussian distribution of mean 0 and variance 1. $\Delta t$ is the integration time step, which was set to 1.0 ps. After each integration step, the position of the center of the confining sphere was slightly adjusted so as to coincide with the center of mass of the DNA molecule, in order for the DNA chain to interact as little as possible with the confinement sphere (Joyeux, 2019b). Fig. 1 shows a typical equilibrated conformation obtained with $P = 100$ protein chains and $C = 500$ crowders ($\rho = 0.16$).

Simulations involving the DNA, proteins and crowders were performed as follows. The DNA chain was first equilibrated inside the confinement sphere. The $C$ crowders were then introduced at random, homogenously distributed and non-overlapping positions and the DNA/crowders system was equilibrated again. The $P$ protein chains were finally introduced at random, homogenously distributed and non-overlapping positions and the DNA/crowders/proteins system was equilibrated once more. The production step, which lasted at least 350 ms, then began. Four simulations with different initial conditions and different sets of random numbers were run for each pair of values of $C = 0, 500, 1000, 1200, 1500, 1750$ or $1875$ and $P = 0, 50, 100, 200$ or $400$. As is illustrated in Fig. 3, equilibrated conformations span a wide range of geometries, ranging from little compacted DNA for low values of $C$ and/or $P$ (middle row of Fig. 3) to rather compact DNA/protein complexes for large values of $C$ and $P$ (bottom row of Fig. 3). Self-associating protein chains always form stable clusters, even for small values of $P$ (top and middle right vignettes of Fig. 3), whereas non-associating protein chains diffuse through the DNA coil (top and middle left vignettes of Fig. 3) and remain close to each other only when the DNA coil is compact (bottom left vignette of Fig. 3).

## 3    Results

### 3.1    DNA compaction laws look different for non-associating proteins and self-associating ones

In order to quantify the size of the DNA coil, its radius of gyration was computed every nanosecond according to

$$R_{\mathrm{g}} = \sqrt{\frac{1}{n}\sum_{k=1}^{n}(\mathbf{r}_k - \mathbf{r}_{\mathrm{CM}}) \cdot (\mathbf{r}_k - \mathbf{r}_{\mathrm{CM}})} \, , \tag{9}$$

where $\mathbf{r}_{\mathrm{CM}}$ denotes the position of the center of mass of the DNA coil, and the mean value $\langle R_{\mathrm{g}}\rangle(\rho, P)$ was obtained by averaging $R_{\mathrm{g}}$ over the four trajectories that were run for each pair of values of $\rho$ and $P$. In particular, in the absence of crowders and proteins, one gets $\langle R_{\mathrm{g}}\rangle(0,0) = 84.9 \pm 1.6$ nm, where the uncertainty $\pm 1.6$ nm denotes the root mean square (RMS) fluctuations of $R_{\mathrm{g}}$ around $\langle R_{\mathrm{g}}\rangle$. The plot of $\langle R_{\mathrm{g}}\rangle(\rho, P)$ as a function of $\rho$ for each value of $P$ is shown in the top plot of Fig. 4 for non-associating proteins and the bottom plot of Fig. 4 for self-associating ones. Comparison of the two plots suggests that DNA compaction laws are different for non-associating proteins and self-associating ones. Indeed, the nearly parallel sets of points in the bottom plot of Fig. 4 indicate that $\langle R_{\mathrm{g}}\rangle$ decreases almost linearly with both $\rho$ and $P$ for self-associating proteins. Stated in other words, $\langle R_{\mathrm{g}}\rangle$ can be reduced by about 1 nm either by increasing $\rho$ by about 0.05 (that is, by adding about 150 crowders) or by adding about 20 self-associating DNA-bridging proteins, whatever the values of $\rho$ and $P$. In contrast, the sets of points in the top plot of Fig. 4 look like rays emerging from a source located close to $(\rho, P) = (0,0)$. This reflects the fact that non-associating proteins are quite inefficient in compacting the DNA coil for low values of $\rho$, whereas their efficiency is larger than that of self-associating proteins for large values of $\rho$ and $P$, because the slope of $\langle R_{\mathrm{g}}\rangle(\rho, P)$ as a function of $\rho$ increases markedly with $P$. For example, about 100 non-associating proteins are needed to reduce $\langle R_{\mathrm{g}}\rangle$ by 1 nm for $\rho = 0$, but this number decreases down to about 10 for $\rho = 0.49$. Stated in other words, non-associating proteins appear to work in synergy with crowders to compact the DNA coil, in the sense that the presence of one component increases the compaction efficiency of the other one, whereas this is not the case for self-associating proteins.

Why the compaction laws displayed in the two plots of Fig. 4 look so different, although the models differ only by the ability of proteins to self-associate or not, is of course an intriguing question. As is discussed in the remainder of this paper, the answer is to be found in the very basic mechanisms which govern the compaction of the DNA coil by DNA-bridging proteins and globular crowders.

## 3.2 $\langle R_{\mathrm{g}}\rangle$ decreases linearly with the number of crowders and the number of protein bridges

The two mechanisms which eventually contribute to the compaction of the DNA coil are (i) a segregative phase separation induced by the demixing of the DNA chain and spherical crowders, and (ii) an associative phase separation induced by the cross-linking of the DNA chain by protein chains.

One may therefore expect that the compaction of the DNA coil is a function of the number of crowders and the number of protein bridges, rather than the number of proteins itself. This conjecture is explored in the present section.

As is schematized in Fig. 5, DNA-bridging proteins can adopt four different conformations with respect to the DNA chain, depending on the number of DNA-binding sites which actually bind to the DNA chain at a given time: (i) "free" proteins do not bind at all to the DNA chain, (ii) one DNA-binding site of "dangling" proteins binds to the DNA chain, whereas the second one does not, (iii) the two DNA-binding sites of "cis-bound" proteins bind to neighboring beads of the DNA chain, and (iv) the two DNA-binding sites of "bridging" proteins bind to beads of the DNA chain which are widely separated in terms of the curvilinear coordinate along the chain. Here, we are interested in "bridging" proteins, and more precisely in the evolution of their mean number $\langle N_{\mathrm{B}} \rangle$ as a function of $\rho$ and $P$. For this purpose, it was considered that one DNA-binding site of a protein chain binds to the DNA chain if it has an interaction energy larger than $2.5\ k_{\mathrm{B}}T$ with at least one DNA bead. Moreover, protein chains must bind to DNA beads whose respective indexes $k$ and $K$ satisfy $|k - K| > 10$ to be considered as "bridging" ones instead of "cis-binding" ones. These thresholds are somewhat arbitrary, but the results discussed below do not depend sensitively thereon (remember that the depth of the interaction between a protein DNA-binding bead and a DNA bead is $3.9\ k_{\mathrm{B}}T$ and that the equilibrium distance between beads $m = 2$ and 6 of a protein chain corresponds to $|k - K| = 4$ (Joyeux, 2021)).

The plot of $\langle N_{\mathrm{B}} \rangle(\rho, P)$ as a function of $\rho$ for each value of $P$ is shown in the top plot of Fig. 6 for non-associating proteins and the bottom plot of Fig. 6 for self-associating ones. For both types of proteins, the evolution of $\langle N_{\mathrm{B}} \rangle$ with $\rho$ and $P$ is well approximated by the empirical function

$$\langle N_{\mathrm{B}} \rangle(\rho, P) = (A + B\rho)P, \tag{10}$$

where a least squares adjustment of the parameters leads to $A = 0.054 \pm 0.011$ nm$^{-1}$ and $B = 0.963 \pm 0.024$ nm$^{-1}$ for non-associating proteins, and $A = 0.442 \pm 0.003$ nm$^{-1}$ and $B = 0.206 \pm 0.007$ nm$^{-1}$ for self-associating ones. The fitted functions are shown as dot-dashed lines in Fig. 6. It is worth noting that the fits are of rather good quality, in the sense that the differences between the values of $\langle N_{\mathrm{B}} \rangle$ obtained from the simulations and from Eq. (10) are small for all points.

The resemblance of the plots in Figs. 4 and 6 suggests that the decrease of the mean radius of gyration of the DNA coil,

$$\Delta\langle R_{\mathrm{g}}\rangle(\rho, P) = \langle R_{\mathrm{g}}\rangle(0,0) - \langle R_{\mathrm{g}}\rangle(\rho, P) \,, \tag{11}$$

may perhaps be a simple linear function of both $\rho$ and $\langle N_{\mathrm{B}}\rangle$

$$\Delta\langle R_{\mathrm{g}}\rangle(\rho, P) = \alpha\rho + \beta\langle N_{\mathrm{B}}\rangle(\rho, P) \,, \tag{12}$$

where $\langle N_{\mathrm{B}}\rangle(\rho, P)$ is taken from Eq. (10). The least squares adjustment of the parameters leads to $\alpha = 19.8 \pm 3.2$ nm$^{-1}$ and $\beta = 0.173 \pm 0.009$ nm$^{-1}$ for non-associating proteins, and $\alpha = 16.8 \pm 1.3$ nm$^{-1}$ and $\beta = 0.103 \pm 0.004$ nm$^{-1}$ for self-associating ones. The points with $\rho = 0.61$ were not included in the fit for non-associating proteins, because they clearly deviate from a straight line in Fig. 4. The result of the fit is shown in the top plot of Fig. 7 for non-associating proteins and the bottom plot of Fig. 7 for self-associating ones. In these figures, symbols represent the values of $\Delta\langle R_{\mathrm{g}}\rangle$ obtained from the simulations and dot-dashed lines those obtained from Eq. (12). It is seen that the simple linear expression in Eq. (12) indeed reproduces reasonably well all the results obtained from simulations, except those involving non-associating proteins at an effective volume fraction of crowders close to the jamming threshold ($\rho = 0.61$).

In spite of the striking difference between the two plots of Fig. 4, $\langle R_{\mathrm{g}}\rangle$ decreases consequently linearly with $\rho$ and $\langle N_{\mathrm{B}}\rangle$ for both non-associating proteins and self-associating ones (Eq. (12)). Moreover, for both protein models, $\langle N_{\mathrm{B}}\rangle$ can be expressed as a simple function of $\rho$ and $P$ (Eq. (10)). Finally, both models share similar values of the parameter $\alpha$ in Eq. (12). This latter point was of course expected, because the term $\alpha\rho$ describes the decrease of $\langle R_{\mathrm{g}}\rangle$ in the absence of proteins of any type ($P = 0$). The consequence of these strong similarities between the two models is that the difference in their DNA compaction capabilities, which is clearly seen in Fig. 4, results uniquely from the differences in parameters $A$ and $B$ of Eq. (10), which describe the evolution of $\langle N_{\mathrm{B}}\rangle$ with $\rho$ and $P$, and parameter $\beta$ of Eq. (12), which describes the evolution of $\langle R_{\mathrm{g}}\rangle$ with $\langle N_{\mathrm{B}}\rangle$. These differences will be rationalized in the forthcoming sections, in order to bring out the mechanisms which govern the organization of the bacterial nucleoid by crowders and DNA-bridging proteins.

### 3.3 Evolution of $\langle N_{\mathrm{B}}\rangle$ with $\rho$ and $P$

As can be checked in Fig. 6, non-associating proteins form a significantly smaller number of DNA bridges than self-associating ones in the absence of crowders ($\rho = 0$). This is reflected in the fact that parameter $A$ of Eq. (10) is approximately 8 times smaller for non-associating proteins than for self-associating ones. However, the number of bridges increases more rapidly as a function of $\rho$ for non-associating proteins than for self-associating ones. This is reflected in the fact that parameter $B$ of Eq. (10) is approximately 5 times larger for non-associating proteins than for self-associating ones. As a consequence, for large effective volume fractions of crowders ($\rho \geq 0.51$) the number of DNA bridges becomes larger for non-associating proteins than for self-associating ones. The purpose of this section is to understand the reasons for these different behaviors.

As was analyzed in detail in (Joyeux, 2017) and (Joyeux, 2018), insertion of spherical crowders at random, non-overlapping and homogeneously distributed positions inside the confinement sphere results in the demixing of the DNA chain from the crowders, with the DNA beads localizing preferentially in the central part of the sphere. This demixing is illustrated in Fig. 8, which shows the enrichment in DNA beads, $Q_{\mathrm{DNA}}$, as a function of the distance $r$ from the center of the confinement sphere, for different values of the effective volume fraction of crowders $\rho$ and in the absence of proteins ($P = 0$). $Q_{\mathrm{DNA}}(r)$ is defined according to

$$Q_{\mathrm{DNA}}(r) = \frac{\langle n(r) \rangle}{n} \frac{R_0^3}{3r^2 dr}, \tag{13}$$

where $\langle n(r) \rangle$ denotes the average number of DNA beads whose center is located at a distance from the center of the sphere comprised between $r$ and $r + dr$. $Q_{\mathrm{DNA}}(r)$ is constant and equal to 1 when the density of DNA beads is uniform in the whole sphere. Fig. 8 confirms that increasing the effective volume fraction of crowders indeed results in a progressive enrichment in DNA beads at the center of the confinement sphere. In turn, this enrichment has the consequence that the average distance from a DNA bead to its nearest-neighbor, $\langle D_{\mathrm{NN}} \rangle$, decreases markedly with increasing $\rho$. This is illustrated in Fig. 9, which shows the evolution of $\langle D_{\mathrm{NN}} \rangle$ with $\rho$ for DNA beads located in the central part of the confinement sphere ($r < 60$ nm), as well as those located in the outer shell ($r \geq 60$ nm). As for DNA bridges in Fig. 6, only those pairs of DNA beads whose indexes $k$ and $K$ satisfy $|k - K| > 10$ were taken into account for estimating $\langle D_{\mathrm{NN}} \rangle$. It is seen in Fig. 9 that the average distance to nearest-neighbors in the central part of the sphere decreases from more than 15 nm for $\rho = 0$ down to about 11 nm for $\rho \geq 0.57$. In the outer shell, this distance is still larger than 14 nm for $\rho = 0.61$. For comparison, the distance between the centers of the two DNA-binding beads ($m = 1$ and 7) of a fully stretched protein chain is $6l_0 = 15.0$ nm. However, because of thermal noise and the flexibility of protein chains, this distance reduces to $11.7 \pm 1.6$ nm when measured in simulations involving only non-associating protein chains. This range is shown as dashed and dot-dashed horizontal lines in Fig. 9. It is seen in this figure that the average distance between the centers of the DNA-binding beads of a protein chain becomes equal to the average distance between the centers of DNA nearest-neighbors in the central part of the confinement sphere only for $\rho \approx 0.5$, that is precisely for the effective volume fraction of crowders where the numbers of DNA bridges formed by non-associating and self-associating proteins become similar.

The picture that emerges from these observations is consequently the following one. In the absence of crowders ($\rho = 0$), the mean distance between DNA nearest-neighbors is significantly larger than the mean distance between the DNA-binding sites of protein chains. As a consequence, thermal fluctuations and collisions are sufficient to break more or less rapidly the relatively weak DNA bridges that were eventually formed by isolated non-associating proteins (remember that the maximum DNA-protein binding energy is $7.8\,k_BT$). In contrast, even for small values of $P$, self-associating proteins assemble in clusters where 4 to 6 protein chains cooperate for the formation of each DNA bridge (see Fig. S8 of (Joyeux, 2021)). Once formed, such bridges are stable and can hardly be destroyed by thermal noise and/or collisions. These considerations explain why self-associating proteins are more efficient than non-associating ones in forming DNA bridges for small values of $\rho$ (and why parameter $A$ of Eq. (10) is larger for self-associating proteins than for non-associating ones). In contrast, for large values of $\rho$, the mean distance between DNA nearest-neighbors becomes equal to the mean distance between the DNA-binding sites of protein chains. This is of little consequence for self-associating protein chains, because the formation of protein clusters ensures that the probability for a protein chain to contribute to a DNA bridge is close to 1, whatever the values of $\rho$ and $\langle D_{NN} \rangle$. In contrast, the decrease of $\langle D_{NN} \rangle$ with increasing values of $\rho$ allows non-associating protein chains to bridge the DNA chain most of the time, in spite of the relative weakness of individual bonds. Indeed, as soon as the DNA bridge formed by a given protein chain is destroyed, the same protein chain reforms another bridge in the immediate neighborhood, because DNA strands are on average at the adequate distance from each other. These considerations explain why the numbers of DNA bridges formed by non-associating and self-associating proteins are similar for large values of $\rho$ (and why parameter $B$ of Eq. (10) is larger for non-associating proteins than for self-associating ones).

Quite interestingly, the validity of the rationale proposed above can be checked to some extent. Indeed, if it is correct, then the number of DNA bridges formed by non-associating proteins depends explicitly on the size of the protein chain. More precisely, the number of DNA bridges must be smaller (and the mean radius of gyration of the DNA coil must be larger) for smaller protein chains. In order to check this point, all simulations with $P = 400$ non-associating protein chains were launched a second time, but with 5-beads protein chains instead of 7-beads ones. This was simply achieved by removing beads $m = 3$ and 5 of the non-associating protein chain shown in Fig. 2, which left the DNA-protein binding strength unchanged. The results obtained with 5-beads protein chains are compared to those for 7-beads protein chains in Fig. 10, where the top and bottom plots show the evolution of $\langle N_B \rangle$ and $\langle R_g \rangle$ with $\rho$, respectively. This figure confirms that $\langle N_B \rangle$ is smaller (and $\langle R_g \rangle$ larger) for 5-beads protein chains than for 7-beads protein chains, which supports the rationale proposed above.

### 3.4   Ability of DNA bridges to compact the DNA coil

After understanding why the evolution with $\rho$ and $P$ of the *number* of DNA bridges is different for non-associating and self-associating proteins, it is important to note that these bridges also differ in their *efficiency* to compact the DNA coil. Indeed, parameter $\beta$ of Eq. (12) is equal to 0.103 for self-associating proteins, which indicates that about 10 DNA bridges are required to decrease $\langle R_g \rangle$ by 1 nm. In contrast, $\beta$ is equal to 0.173 for non-associating proteins, so that only 6 DNA bridges are required to achieve the same goal. The purpose of this section is to understand why DNA bridges formed by non-associating proteins are about 50% more efficient than those formed by self-associating proteins to compact the DNA coil.

The answer to this question is to be found in Fig. 11, which shows the time evolution of the mean squared displacement (MSD) of DNA beads (top plot), self-associating protein beads (middle plot), and non-associating protein beads (bottom plot). Each plot displays several curves obtained for different systems, whose compositions are indicated close to the curves. It is obvious from this figure that non-associating proteins always preserve a large mobility, even when interacting with both the DNA and spherical crowders (note that the saturation at MSD≈0.0035 µm² for the system DNA/400 P/$\rho$ = 0.49 corresponds to the size of the DNA/protein complex). In particular, the mobility of non-associating proteins is always much larger than the mobility of DNA beads when both components are present in the system. Systems containing the DNA chain and non-associating proteins can therefore be described as a deformable DNA coil, which is cross-linked by DNA-bridging proteins, with all proteins preserving a large mobility inside the coil (bottom left vignette of Fig. 3). As a result, the DNA/protein complex is nearly spherical even in the absence of crowders. For example, the mean value of the asphericity coefficient

$$e = \frac{2\lambda_3 - \lambda_1 - \lambda_2}{2\lambda_3},$$

(14)

where $0 \leq \lambda_1 \leq \lambda_2 \leq \lambda_3$ are the three eigenvalues of the matrix of inertia of the proteins, is equal to 0.08 for $P = 400$ and $\rho = 0$ (remember that $e = 0$ for a sphere, whereas $e = 1/2$ for a flat surface).

In contrast, the mobility of self-associating proteins is always small, even in the absence of DNA and crowders. This is, of course, due to the fact that all proteins assemble rapidly into one stable cluster (top right vignette of Fig. 3) and that the mobility of individual proteins reflects henceforth the mobility of the cluster. In particular, the mobility of self-associating proteins is always much smaller than the mobility of DNA beads when both components are present in the system. Systems containing the DNA chain and self-associating proteins can therefore be described as a stable network of proteins, with the DNA chain wrapping around it (bottom right vignette of Fig. 3). In the absence of crowders, the protein network is more planar than spherical, with a mean value of the asphericity coefficient $e$ equal to 0.35 for $P = 400$ and $\rho = 0$.

The picture that emerges from these observations is consequently that the compaction of the DNA coil by non-associating proteins is a more dynamical process than the compaction by self-associating ones. For self-associating proteins, the DNA chain essentially strives to maximize its total binding energy with the little deformable network of proteins. Stated in other words, the DNA/protein complex looks like a gel, whose shape is not necessarily spherical. In contrast, mobile non-associating proteins adapt rapidly to the new conformation of the DNA coil as it itself compacts under the influence of protein bridges. Bridges that become less efficient in compacting the DNA coil are continuously replaced by new bridges that are more efficient with this respect, so that the DNA coil compacts further. For large numbers of non-associating proteins and crowders, one eventually obtains the compact spherical DNA/protein complex shown in the bottom left vignette of Fig. 3. These are the reasons, why parameter $\beta$ is significantly larger for non-associating proteins than for self-associating ones.

## 4    Discussion

The simulation work reported in the present paper therefore supports the conclusion that DNA-bridging proteins and globular crowders cooperate in compacting the DNA coil in the whole range of biologically relevant concentrations of these components, in the sense that the radius of gyration of the DNA coil decreases linearly with both the number of globular crowders and the number of cross-links formed by DNA-bridging proteins. Moreover, this work suggests that self-associating proteins are more efficient than non-associating ones in bridging the DNA coil as long as the average distance between DNA strands is larger than the length of DNA-bridging proteins, that is, for small enough values of the effective volume ratio of globular crowders. Instead, when the average distance between DNA strands becomes shorter than the length of DNA-bridging proteins, then the higher mobility of non-associating proteins enables them to cross-link and compact the DNA coil more efficiently than self-associating ones.

These conclusions transpose probably safely to real life, because the coarse-grained model reproduces satisfactorily the corresponding relevant features of bacterial cells:

- the concentration in DNA base pairs in *E. coli* cells is approximately 4.6 millions per 1.3 μm$^3$, that is, ≈6 mM, against ≈5 mM for the model. The average distance between DNA strands is consequently correctly reproduced by the model.

- it is known that H-NS monomers are composed of 136 amino acids and bind to the DNA via their C-terminal domain (residues 89-136) (Shindo et al., 1995), whereas their N-terminal domain (residues 1-63) forms stable antiparallel dimers (Bloch et al., 2003; Arold et al., 2010). The two moieties are connected by a flexible linker (residues 64-88), which also participates in the oligomerization of the dimers (Smyth et al., 2000; Badaut et al., 2002; Arold et al. 2010). A recent X-ray structure (PDB 3NR7) has shown that the total length of two antiparallel N-terminal domains is 12.1 nm (Arold et al. 2010). H-NS *dimers* are consequently adequately modeled by protein chains composed of 7 beads, with DNA-binding beads $m = 1$ and 7 rotating freely around oligomerization beads $m = 2$ and 6, which are separated at equilibrium by a distance $4l_0 = 10.0$ nm.

- it has been determined experimentally that the enthalpy change upon binding of a H-NS molecule to a 16 bp site on the DNA molecule is of the order of $10.0\ k_\mathrm{B}T$ at 298 K (Ono et al., 2005), which correlates well with model protein chains binding to two consecutive beads of the DNA chain with a total energy of $7.8\ k_\mathrm{B}T$.

- finally, the number of H-NS and StpA monomers per complete genomic DNA (4.6 Mbp) appears to be relatively stable over all cell cycle steps, being of the order of 8000 H-NS monomers and 10000 StpA monomers (Azam et al., 1999), which are uniformly distributed in the whole nucleoid (Azam et al., 2000). Owing to the $\approx 200$ reduction factor of the model compared to real cells and to the fact that each DNA bridge is composed of two monomers, the effects of these nucleoid proteins are best modeled by introducing $P = 50$ protein chains in the simulations. Still, simulations with larger values of $P$ were also launched because it is known that, in addition to H-NS and StpA, which are the principal bridgers of the DNA coil, a few other nucleoid proteins, such as Lrp, SMC and Fis, also display some DNA bridging capabilities (Luijsterburg et al., 2006; Luijsterburg et al., 2008).

Quite interestingly, some of the predictions of the coarse-grained model may eventually be checked experimentally in the future, because mutants of H-NS with a reduced capacity to form oligomers (Spurio et al., 1992; Spurio et al., 1997) or displaying instead enhanced oligomerization properties in solution (Giangrossi et al., 2014) have been proposed. On the other hand, the present work has several limitations, whose lifting requires additional work:

- several experiments suggest that the degree of oligomerization of H-NS proteins in solutions that contain only the proteins increases with the concentration of monomers and that tetramers and hexamers may predominate in the 30-100 μM range (Smyth et al., 2000; Ceschini et al., 2000). The coarse-grained model reproduces correctly the experimentally determined enthalpy change of $\approx 10.0\, k_{\mathrm{B}}T$ when two H-NS dimers bind and form a tetramer (Ceschini et al., 2000), as well as the fact that larger oligomers predominate over dimers at physiological concentrations of DNA-bridging proteins (Smyth et al., 2000; Ceschini et al., 2000). However, simulations systematically lead to a few large clusters composed of tens to hundreds of protein chains, which are probably much too large compared to observed oligomers. This opens the interesting possibility that in living cells the size of protein oligomers may be large enough to compensate for the short length of the dimers compared to the average distance between DNA strands when the DNA coil is decompacted, but small enough not to alter their mobility, which facilitates further compaction of the DNA coil when it has already been compacted to some extent.

- supercoiling has not been taken into account in the present work, although it does contribute to the compaction of the DNA coil. A first reason is that it would not have been practically feasible to span the whole range of values of $P$ and $\rho$ and in addition vary the value of DNA's superhelical density. A second, more fundamental reason is that supercoiling probably interacts non-trivially with both globular crowders and DNA-bridging proteins. Indeed, it has been shown that the effects of supercoiling and demixing are additive only in a limited range of biologically relevant values of the parameters, because the plectonemes generated by supercoiling are stiffer and less easily compacted by globular macromolecules than torsionally relaxed DNA (Joyeux, 2019). Moreover, interactions of plectonemes and DNA-bridging proteins are most probably also rather complex but have not been addressed so far. Therefore, systems composed of supercoiled DNA and DNA-bridging proteins must first be studied in some detail before the full system composed of supercoiled DNA, globular crowders and DNA-bridging proteins is investigated for selected values of the parameters.

- the model is meant to reproduce a bacterial cell at a scale of about 1/200 with realistic concentrations of DNA and DNA-bridging proteins. It has been implicitly assumed throughout the paper that the conclusions drawn from the model extend readily to a cell whose volume is 200 times larger. This is probably the case for the compaction of the DNA coil by DNA-bridging proteins, because compaction results here from the minimization of the total enthalpy of the system, which increases almost linearly with its size. In contrast, compaction of the DNA coil by globular crowders results from the maximization of the total entropy of the system, which does not necessarily increase linearly with its size. This uncertainty has been circumvented to some extent by investigating the full range of concentrations of globular crowders (from zero concentration to jamming), and consequently a broad range of crowders-induced compaction of the DNA coil, but this is a point which certainly deserves further work.

To conclude, let us mention briefly that Liquid-Liquid Phase Separation (LLPS) is being increasingly evoked as an important mechanism for the organization of the cytoplasm, because it could provide a rationale for the creation of organelles without bounding membrane (Hyman et al., 2014; Berry et al., 2015; Brangwinne et al., 2015). Most examples investigated to date which evoke LLPS as a possible building mechanism, like the formation of the nucleolus (Boisvert et al., 2007), the centrosome (Mahen and Venkitaraman, 2012) and stress granules (Buchan and Parker, 2009), would involve associative phase separation, where two or more components selectively attract each other and form regions that are enriched in these components. These regions usually have the form of a droplet and indeed behave like liquid droplets immersed in another liquid, but it is at present not completely clear that they result from a true phase separation in the thermodynamic sense (if only because these regions are of finite size). Still, keeping with the idea of liquid droplets, it may be worth emphasizing that many simulations involving non-associating proteins, namely those with $P = 100$ and $\rho \geq 0.49$, $P = 200$ and $\rho \geq 0.39$, and $P = 400$ and $\rho \geq 0.16$, lead to DNA/protein complexes which also look like liquid droplets. Indeed, they are nearly spherical (bottom left vignette of Fig. 3), protein chains move rapidly throughout the complex (bottom plot of Fig. 11), and the exchange of molecules through the "surface" of the complex is easy. For certain concentrations of globular crowders and DNA-bridging proteins, the nucleoid itself consequently looks like a liquid droplet. However, the important difference between the examples cited above and the nucleoid is that the formation of the nucleoid would involve both an associative phase separation (proteins tend to aggregate with the DNA) and a dissociative one (globular crowders tend to demix from both the DNA and the proteins). As a consequence, the size of the nucleoid and the concentration of proteins inside the nucleoid can be easily varied by adjusting the effective volume ratio of the crowders at constant protein number. Stated in other words, when considering only the components of the coarse-grained model, the nucleoid would behave like a *compressible liquid droplet* composed of DNA and DNA-bridging proteins and immersed in a dense gas of crowders. This analogy may deserve further attention and investigation, although real liquids are of course usually only little compressible.

## 5    Conflict of Interest

The authors declare that the research was conducted in the absence of any commercial or financial relationships that could be construed as a potential conflict of interest.

## 6    Funding

This work was supported by the research funds CNRS MITI Modélisation du Vivant 2021 and 2022 "StatPhysProk".

## 8    Figure captions

**Figure 1**. Snapshot extracted from a simulation with $P = 100$ non-associating protein chains and $C = 500$ crowders ($\rho = 0.16$). The long chain with white beads represents the circular DNA molecule, whereas shorter linear chains with blue and red beads represent non-associating proteins (red beads denote the DNA-binding sites of each protein chain). Crowders are shown as yellow balls. One half of the confinement sphere and of the crowders have been removed, in order for the DNA and the proteins to be seen more clearly.

**Figure 2**. Diagram of self-associating and non-associating protein chains. Index $m$ is indicated for each bead. Red disks ($m = 1$ and 7) represent DNA-binding beads, which rotate freely around beads $m = 2$ and 6, respectively. For self-associating protein chains, green disks ($m = 2$ and 6) represent oligomerization beads, which can bind to beads $m = 2$ and 6 of other self-associating protein chains. Blue disks represent protein beads which repel all other components of the system.

**Figure 3**. Snapshots extracted from simulations with $P = 50$ protein chains and $C = 500$ crowders, corresponding to $\rho = 0.16$ (top and middle row), or $P = 400$ protein chains and $C = 1500$ crowders, corresponding to $\rho = 0.49$ (bottom row). The left column shows results obtained with non-associating protein chains, whereas the right column shows the corresponding results obtained with self-associating protein chains. Vignettes in the top row zoom in on the central parts of the vignettes in the middle row. The circular DNA chain is shown as a semi-transparent white tube and linear protein chains as blue/green/red tubes, where red parts correspond to DNA-binding sites and green ones to oligomerization sites. Crowders are not shown. The confinement sphere is shown as a brown circle.

**Figure 4**. Plot of the mean value of the radius of gyration of the DNA coil, $\langle R_g \rangle$, as a function of the effective volume fraction of crowders, $\rho$, for different values of the number $P$ of non-associating proteins (top plot) or self-associating proteins (bottom plot). Different symbols and colors are used for different values of $P$. Error bars represent the RMS fluctuations of $R_g$ around $\langle R_g \rangle$.

**Figure 5**. Diagram showing the different modes of interaction between protein chains and the DNA chain. DNA beads are shown in gray and protein beads in red, green or blue, according to the same color code as in Figs 1-3.

**Figure 6**. Plot of the mean number of protein bridges, $\langle N_B \rangle$, as a function of the effective volume fraction of crowders, $\rho$, for different values of the number $P$ of non-associating proteins (top plot) or self-associating proteins (bottom plot). Symbols represent the values obtained from simulations. The same colors and symbols as in Fig. 4 are used to discriminate against different values of $P$. Error bars represent the RMS fluctuations of $N_B$ around $\langle N_B \rangle$. Dot-dashed lines show the result of the fit of simulation points with Eq. (10).

**Figure 7**. Plot of the decrease of the mean radius of gyration of the DNA coil, $\Delta \langle R_g \rangle$, as a function of the effective volume fraction of crowders, $\rho$, for different values of the number $P$ of non-associating proteins (top plot) or self-associating proteins (bottom plot). Symbols represent the values obtained from simulations. The same colors and symbols as in Figs. 4 and 6 are used to discriminate against different values of $P$. Error bars represent the RMS fluctuations of $R_g$ around $\langle R_g \rangle$. Dot-dashed lines show the result of the fit of simulation points with Eq. (12).

**Figure 8**. Plot of the enrichment in DNA beads, $Q_{\text{DNA}}$, as a function of the distance $r$ from the center of the confinement sphere, for different values of the effective volume fraction of crowders $\rho$. These plots were obtained by averaging the position of DNA beads in simulations without proteins ($P = 0$). Oscillations in the plot of $Q_{\text{DNA}}$ observed for the largest values of $\rho$ reflect the fact that spherical crowders tend to organize in concentric shells at large crowder density.

**Figure 9**. Plot of the average distance between nearest-neighbor DNA beads, $\langle D_{\text{NN}} \rangle$, as a function of the effective volume fraction of crowders, $\rho$, in the central part of the confinement sphere ($r < 60$ nm, + symbols), as well as the outer shell ($r \geq 60$ nm, × symbols). These results were obtained from the same simulations with $P = 0$ as Fig. 8. The horizontal dashed line shows the average distance between the centers of the DNA-binding sites of non-associating protein chains in simulations involving only these chains, and the dot-dashed lines the fluctuations of this distance around the mean value.

**Figure 10**. Plot of $\langle N_{\text{B}} \rangle$ (top) and $\langle R_{\text{g}} \rangle$ (bottom), as a function of the effective volume fraction of crowders, $\rho$, for $P = 400$ non-associating proteins. Values obtained from simulations with 7-beads protein chains are shown as filled lozenges and those obtained with 5-beads protein chains as empty lozenges. Error bars represent the RMS fluctuations of $N_{\text{B}}$ and $R_{\text{g}}$ around their mean values.

**Figure 11**. Time evolution of the mean squared displacement of DNA beads (top plot), self-associating protein beads (middle plot), and non-associating protein beads (bottom plot). Each plot displays several curves, which all correspond to different systems. The composition of each system is indicated close to the curve. Labels "DNA", "400 SAP", "400 P", and "$\rho = 0.49$" are used to indicate that the system contained the DNA chain, 400 self-associating proteins, 400 non-associating proteins, and spherical crowders at an effective volume fraction $\rho = 0.49$, respectively.

**FIGURE 1**

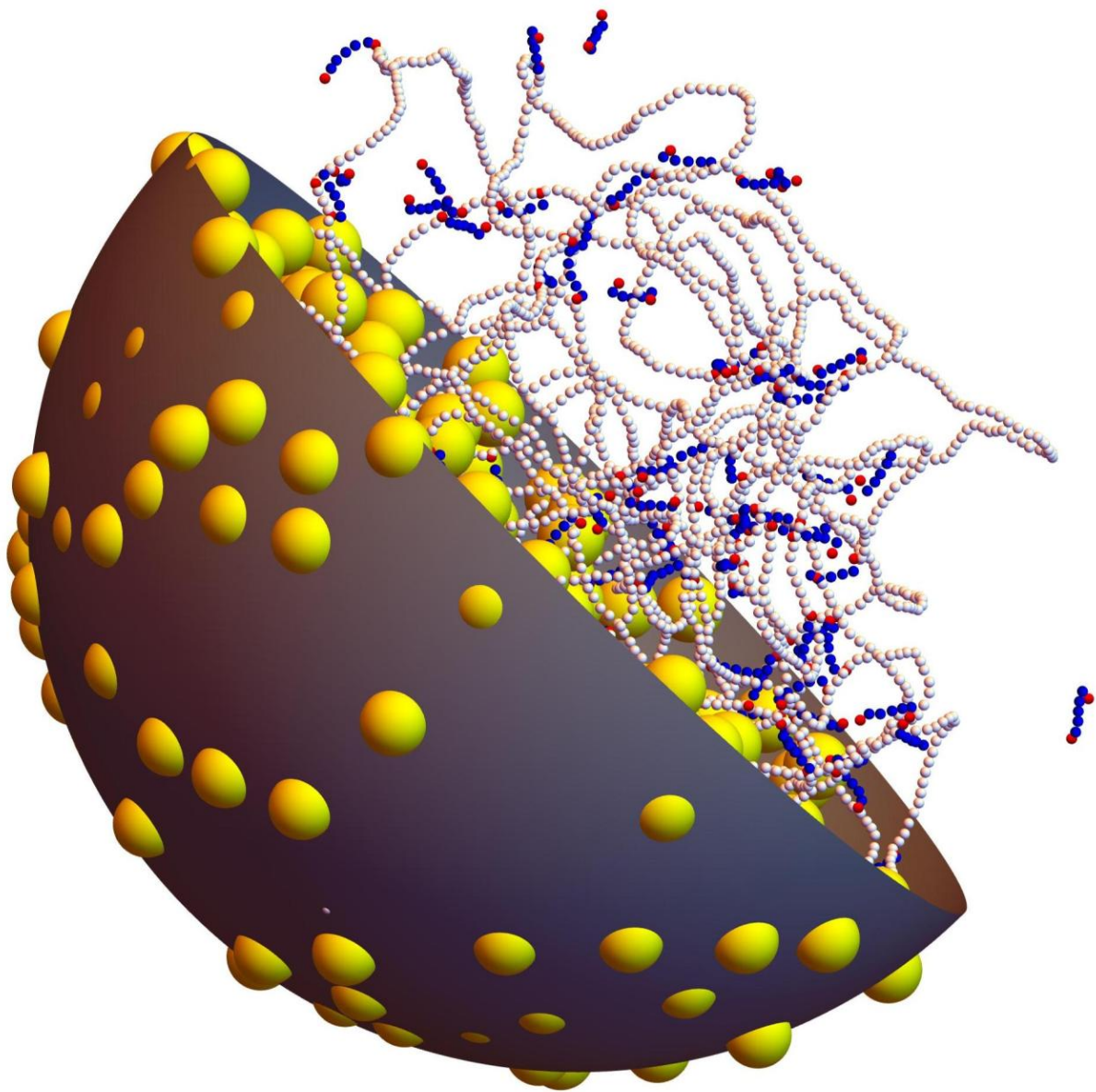

**FIGURE 2**

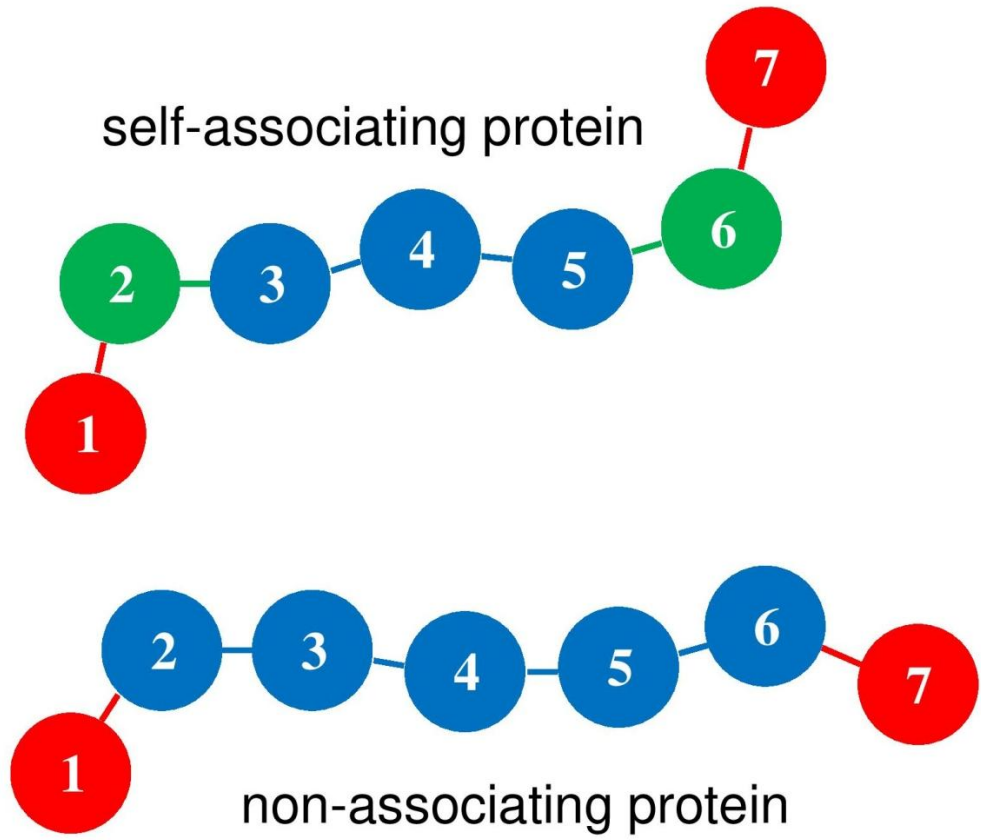

self-associating protein

non-associating protein

**FIGURE 3**

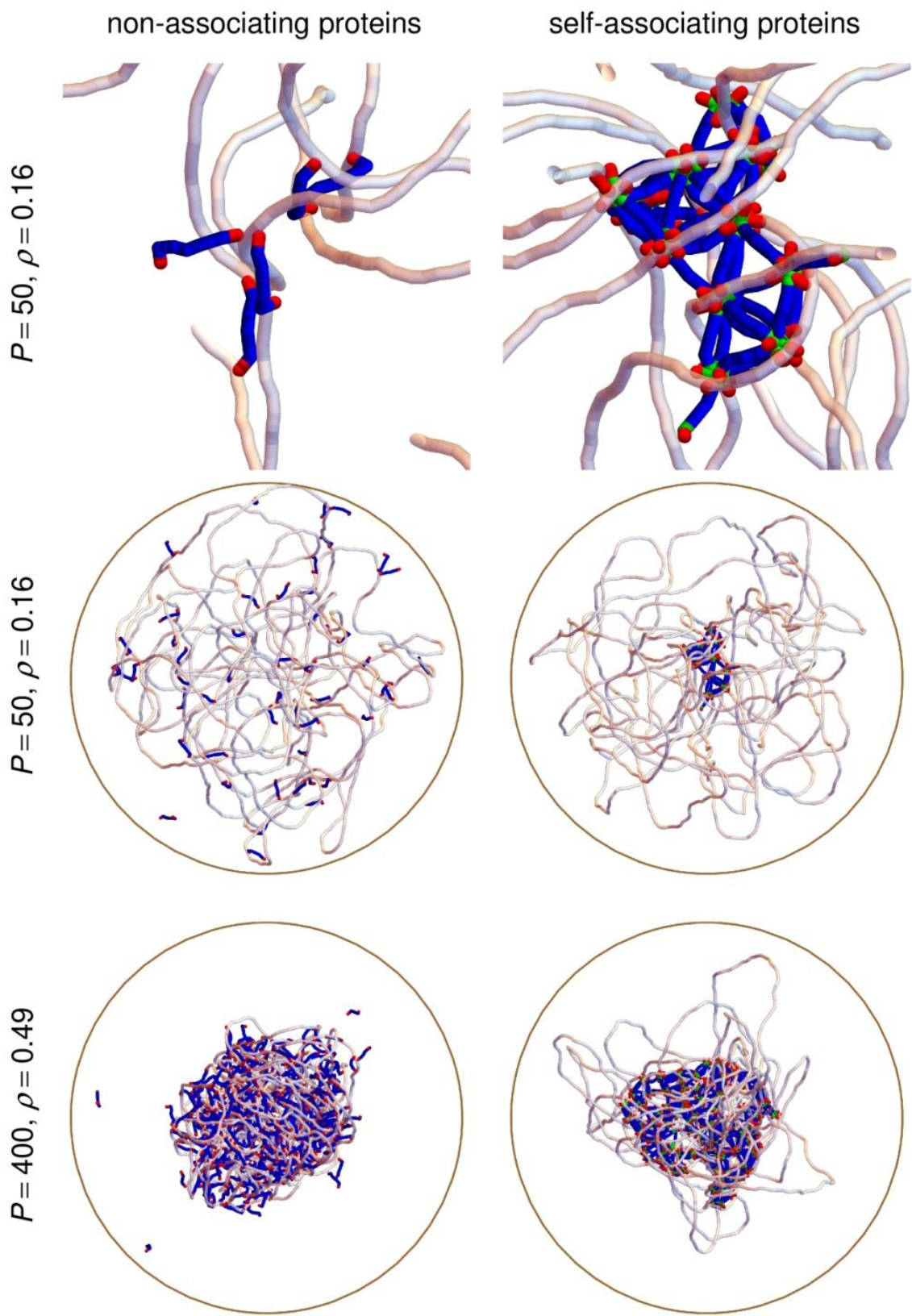

**FIGURE 4**

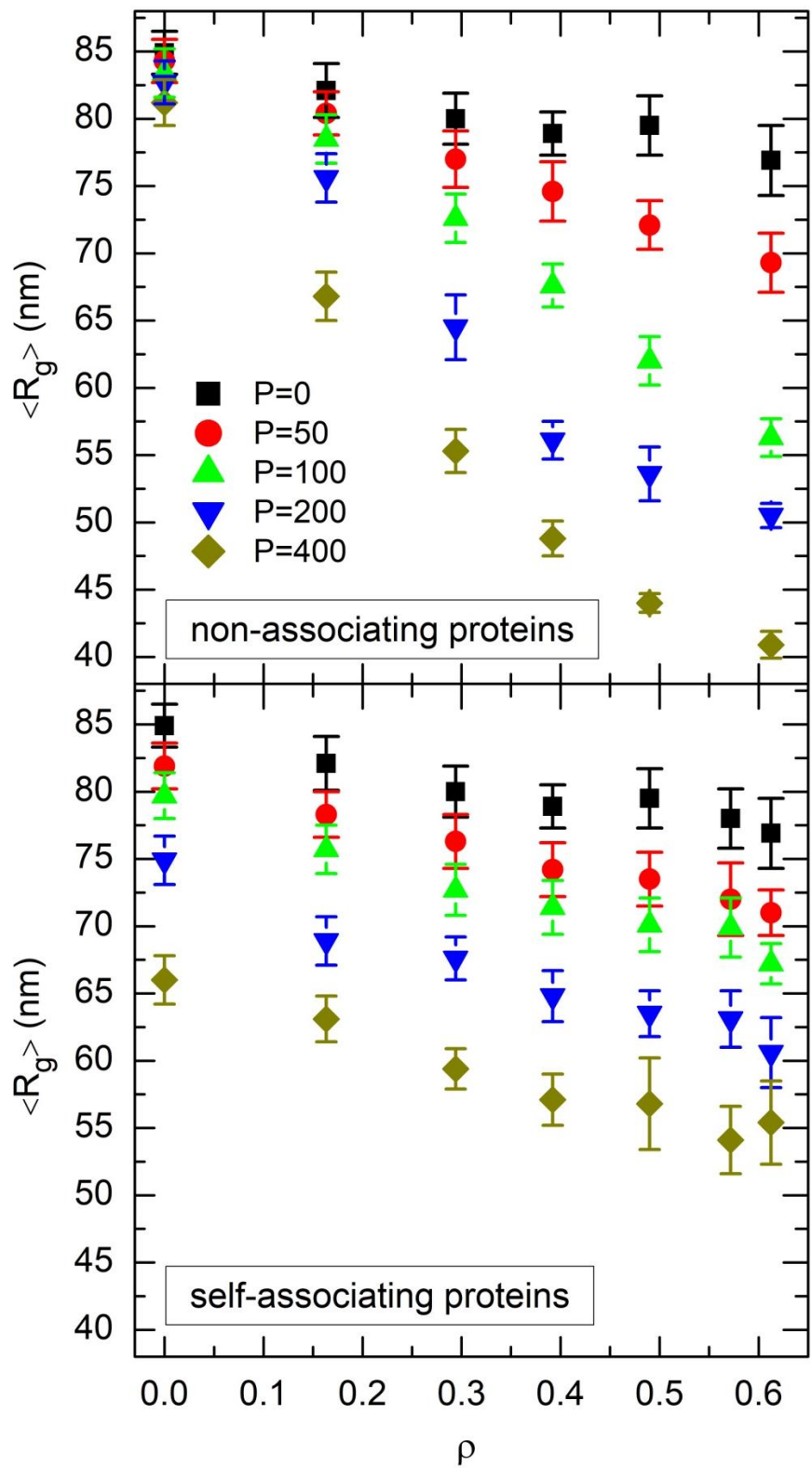

**FIGURE 5**

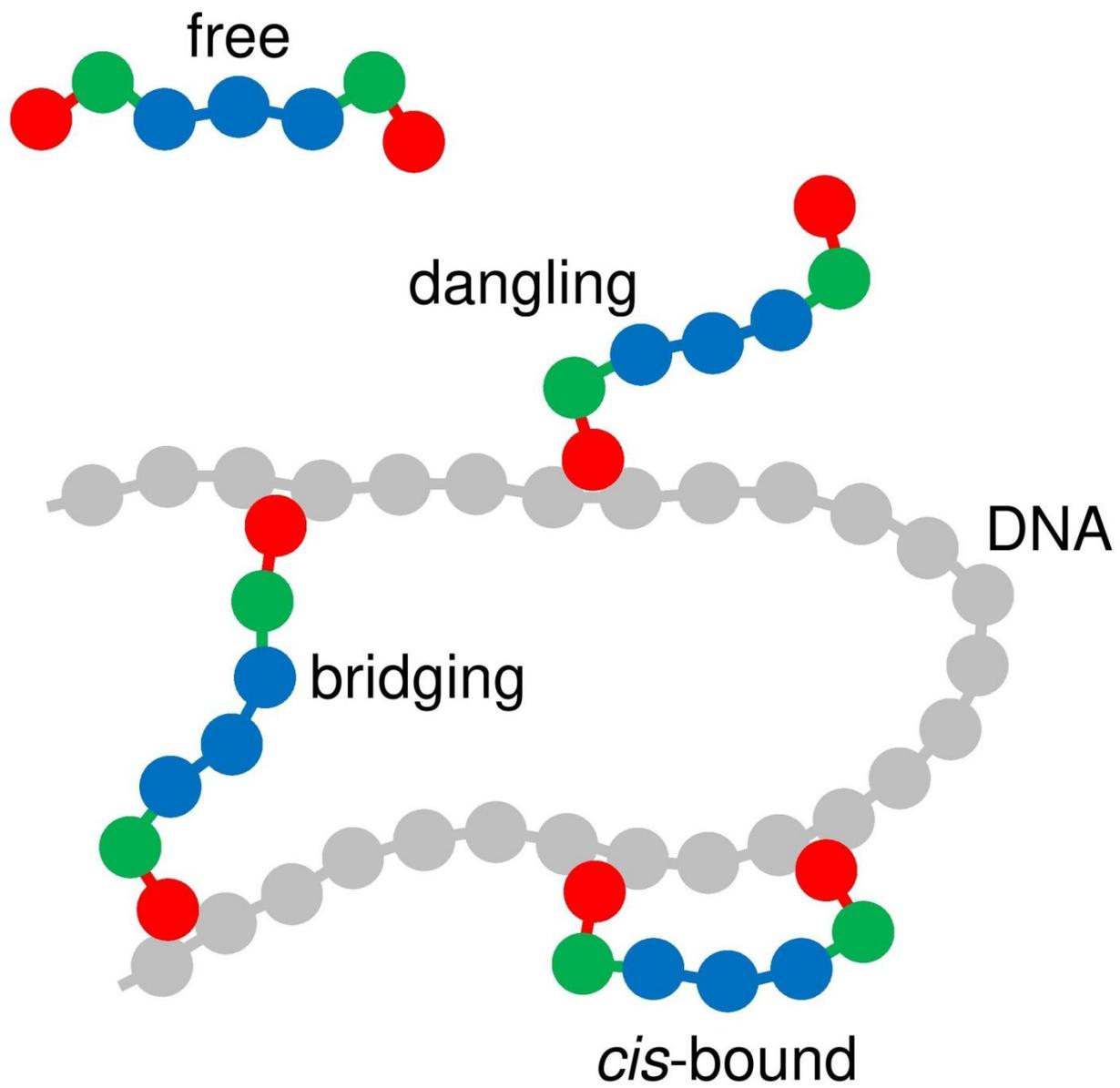

**FIGURE 6**

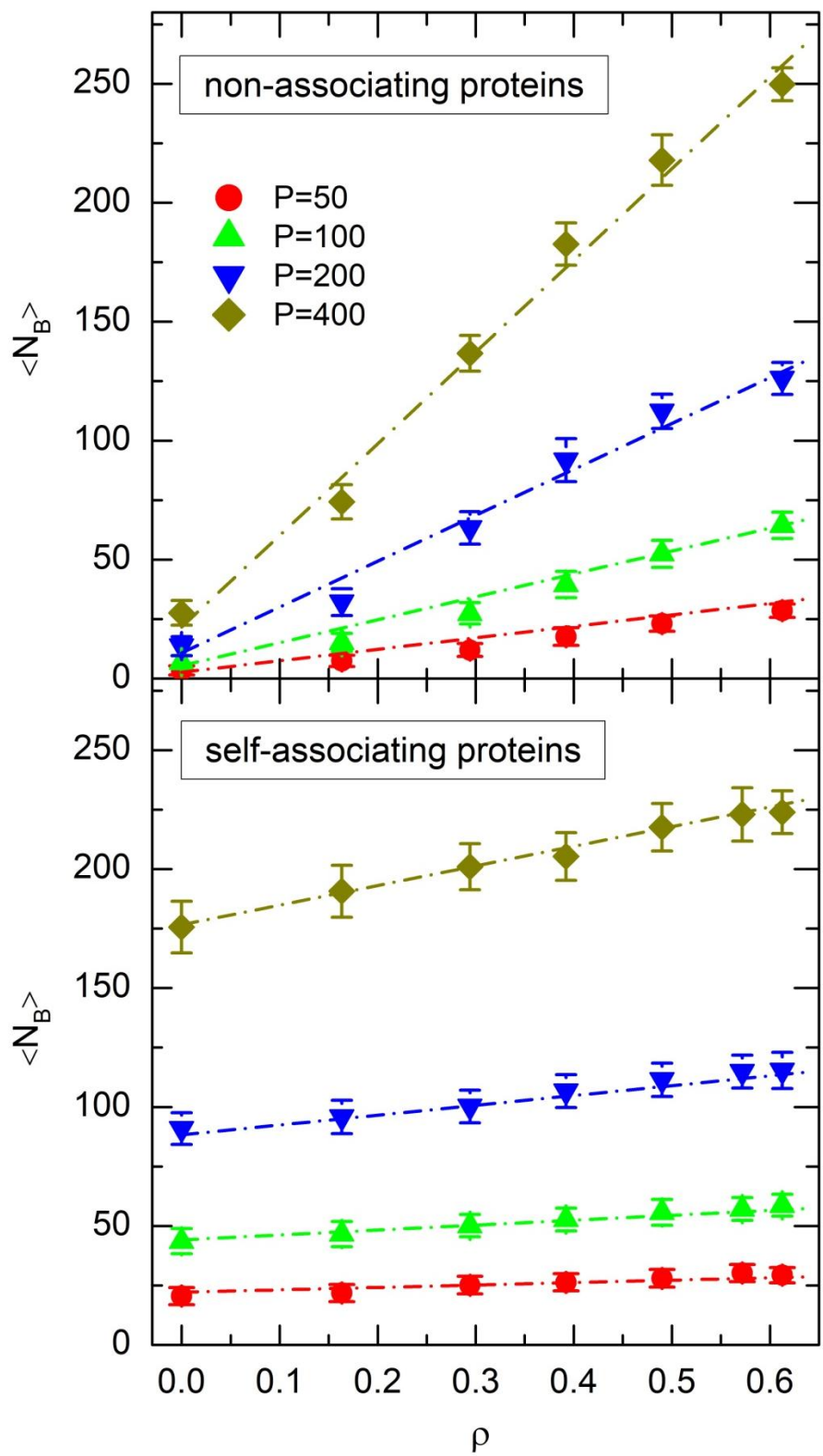

**FIGURE 7**

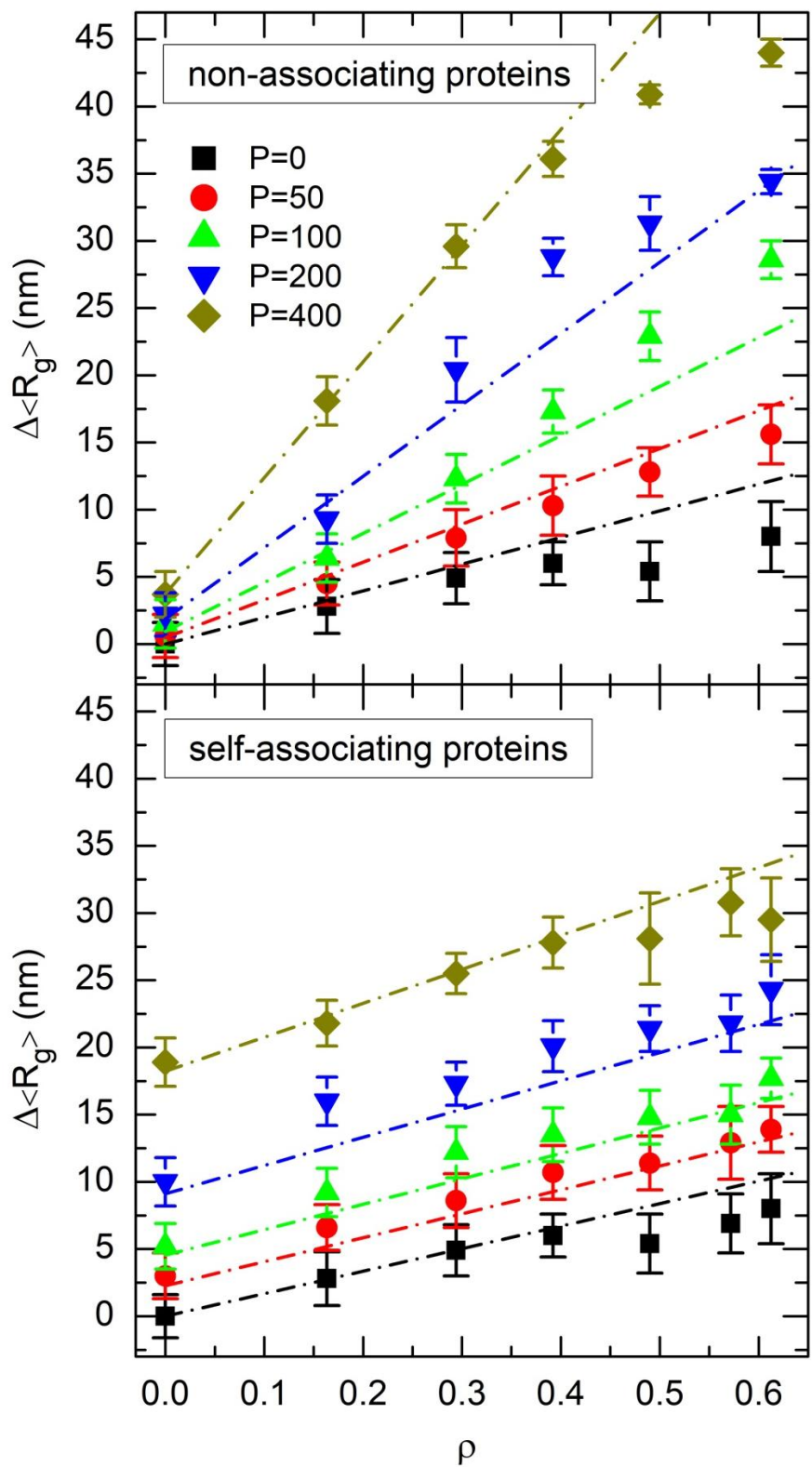

**FIGURE 8**

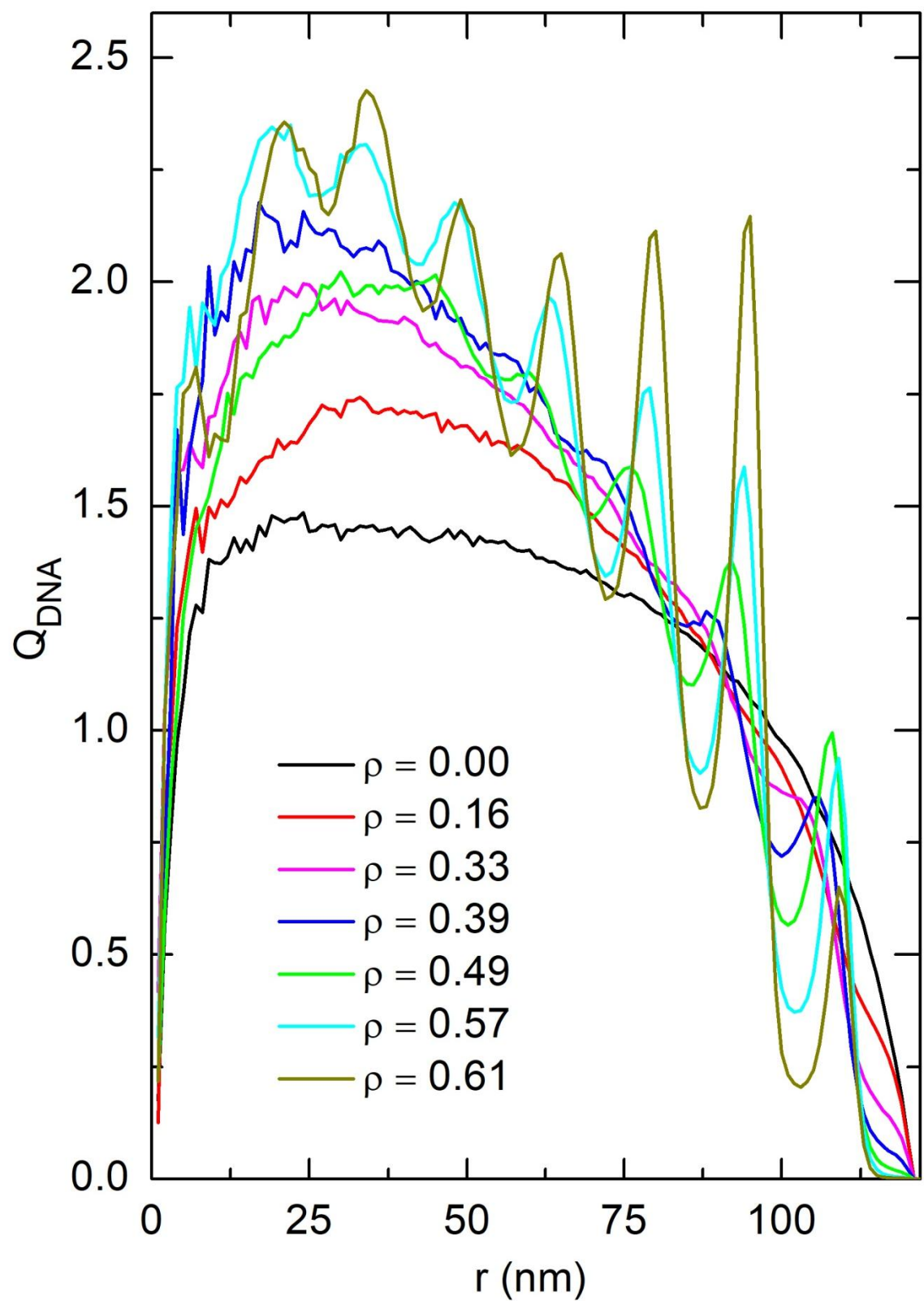

**FIGURE 9**

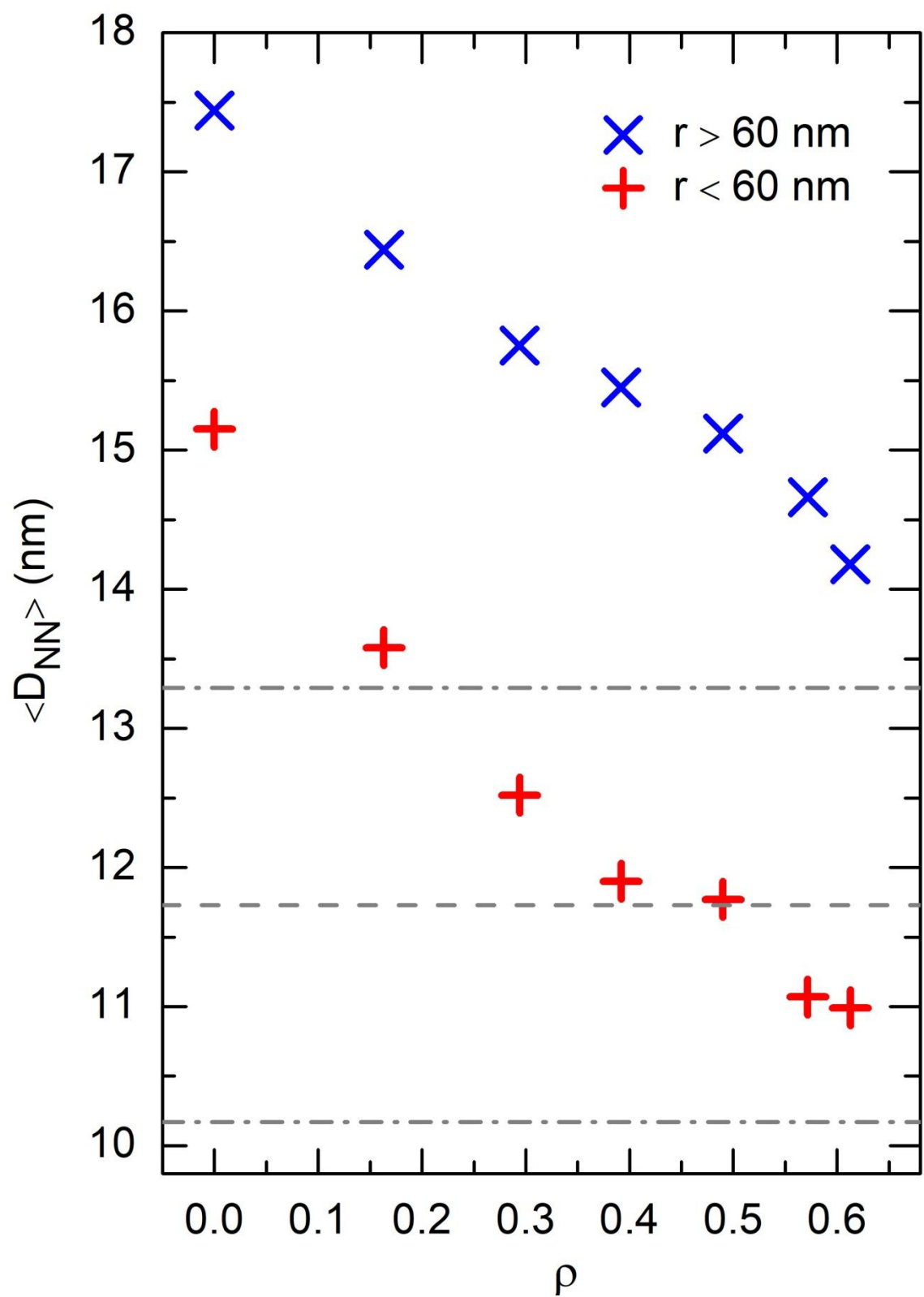

**FIGURE 10**

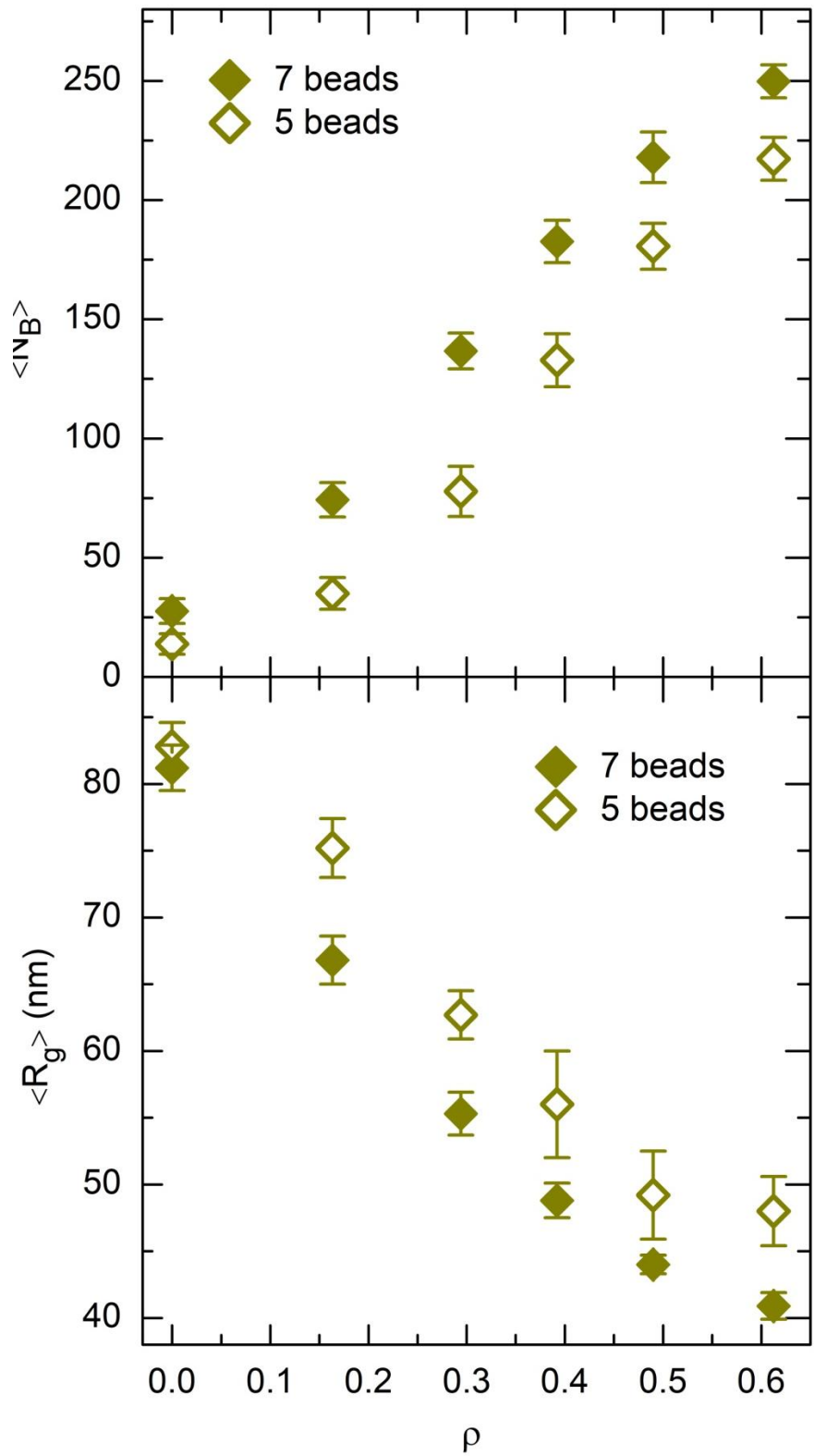

**FIGURE 11**

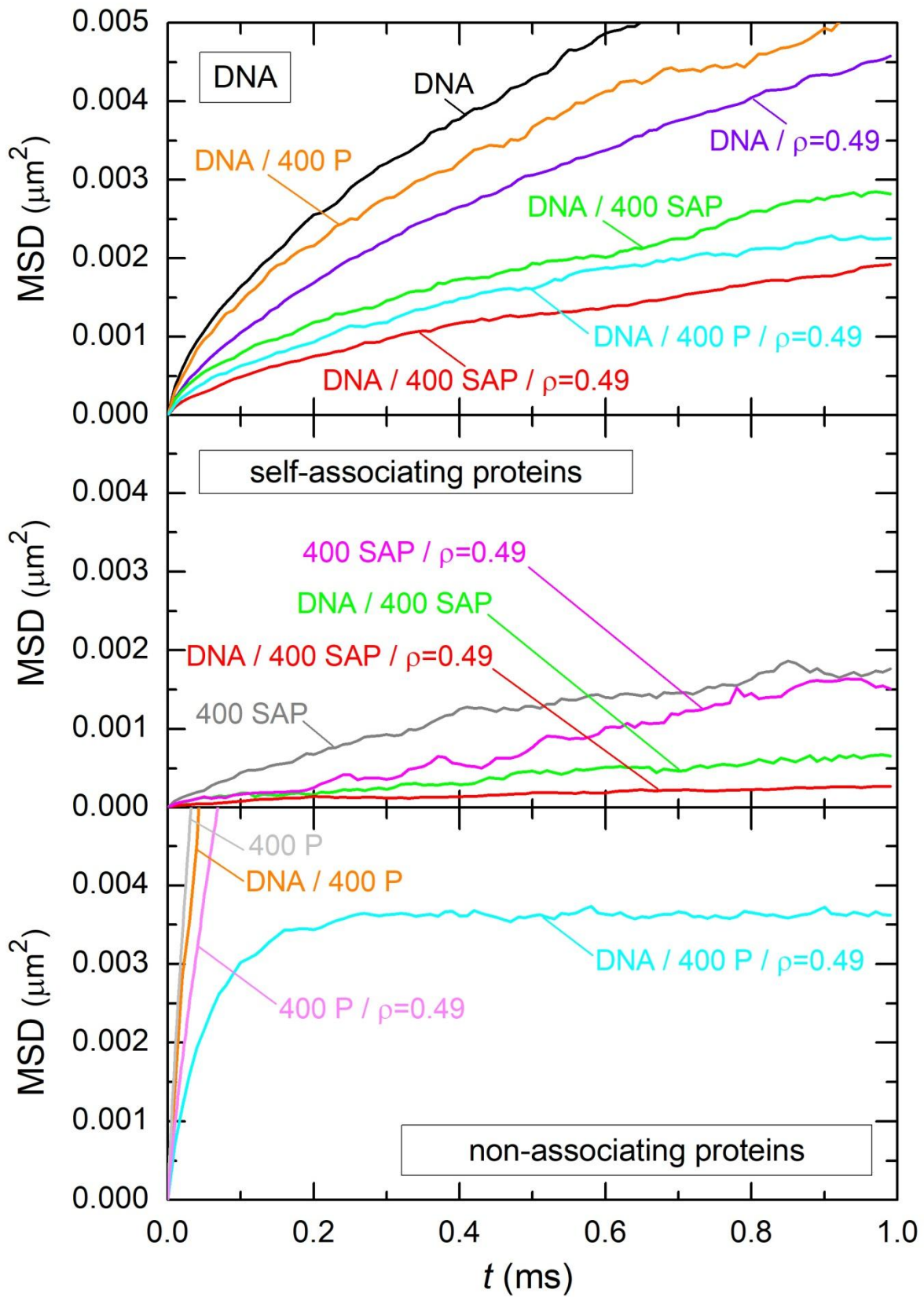